\RequirePackage{fixltx2e}
\RequirePackage{fix-cm}

\documentclass[%
floatfix,
showkeys,
twocolumn, %
nofootinbib,
]{revtex4-1}
\usepackage{cmap}

\usepackage{ucs}
\usepackage[utf8x]{inputenc}
\usepackage[T1,T2A]{fontenc}
\usepackage[german,russian,english]{babel}

\usepackage[sort&compress]{natbib}
\usepackage{amsmath}
\usepackage{amssymb}

\usepackage{mathtools}
\mathtoolsset{
showonlyrefs,
mathic = true
}

\allowdisplaybreaks

\usepackage{hyperref}
\hypersetup{backref,
 colorlinks=false}
\hypersetup{pdfborder=0 0 0}

\usepackage{breakurl}

\usepackage{microtype}
\UseMicrotypeSet[protrusion]{alltext}

\usepackage{fixltx2e}

\usepackage{nicefrac}

\usepackage{physymb}

\makeatletter
\def\ps@pprintTitle{%
     \let\@oddhead\@empty
     \let\@evenhead\@empty
     \let\@oddfoot\@empty
     \let\@evenfoot\@oddfoot}
\makeatother
\usepackage{setspace}

\renewcommand{\d}{\mathrm{d}}

\newcommand{\setR}{\mathbb{R}}
\newcommand{\setZ}{\mathbb{Z}}

\newcommand{\crd}[1]{\underline{\vphantom{j}{#1}}}

\begin{document}
\graphicspath{{image/axiom-sdu/en/}}

\title{One-Step Stochastic Processes Simulation Software Package}

\author{E. G. Eferina}
\email{eg.eferina@gmail.com}
\affiliation{Department of Applied Probability and Informatics\\
Peoples' Friendship University of Russia\\
Miklukho-Maklaya str. 6, Moscow, 117198, Russia}

\author{A. V. Korolkova}
\email{avkorolkova@gmail.com} 
\affiliation{Department of Applied Probability and Informatics\\
Peoples' Friendship University of Russia\\
Miklukho-Maklaya str. 6, Moscow, 117198, Russia}

\author{M. N. Gevorkyan}
\email{mngevorkyan@sci.pfu.edu.ru} 
\affiliation{Department of Applied Probability and Informatics\\
Peoples' Friendship University of Russia\\
Miklukho-Maklaya str. 6, Moscow, 117198, Russia}

\author{D. S. Kulyabov}
\email{yamadharma@gmail.com}
\affiliation{Department of Applied Probability and Informatics\\
Peoples' Friendship University of Russia\\
Miklukho-Maklaya str. 6, Moscow, 117198, Russia}
\affiliation{Laboratory of Information Technologies\\
Joint Institute for Nuclear Research\\
Joliot-Curie 6, Dubna, Moscow region, 141980, Russia}

\author{L. A. Sevastyanov}
\email{leonid.sevast@gmail.com}
\affiliation{Department of Applied Probability and Informatics\\
Peoples' Friendship University of Russia\\
Miklukho-Maklaya str. 6, Moscow, 117198, Russia}
\affiliation{Bogoliubov Laboratory of Theoretical Physics\\
Joint Institute for Nuclear Research\\
Joliot-Curie 6, Dubna, Moscow region, 141980, Russia}

\thanks{Published in:
  E.~G. Eferina, A.~V. Korolkova, M.~N. Gevorkyan, D.~S. Kulyabov, L.~A.
  Sevastyanov, One-Step Stochastic Processes Simulation Software Package,
  Bulletin of Peoples’ Friendship University of Russia. Series ``Mathematics.
  Information Sciences. Physics''~(3) (2014) 46--59.
}

\thanks{Sources:
  \url{https://bitbucket.org/yamadharma/articles-2014-axiom-sdu}}

\begin{abstract}

  \begin{description}

  \item[Background] It is assumed that the introduction of stochastic in
    mathematical model makes it more adequate. But there is
    virtually no methods of coordinated (depended on
    structure of the system) stochastic introduction into deterministic
    models. Authors have improved the method of
    stochastic models construction for the class of one-step processes and
    illustrated by models of population dynamics.
    Population dynamics was chosen for study because
    its deterministic models were sufficiently well explored that
    allows to compare the results with already known ones.

  \item[Purpose] To optimize the models creation
    as much as possible some routine operations should be automated. In this
    case, the process of drawing up the model equations can be
    algorithmized  and implemented in the computer
    algebra system. Furthermore, on the basis of these results a set of programs for numerical experiment
        can be obtained    .

  \item[Method] The computer algebra system Axiom is used for analytical calculations implementation. 
  To perform the numerical
    experiment FORTRAN and Julia languages are used. The method Runge--Kutta method for
    stochastic differential equations is used as numerical method.

  \item[Results] The program compex for creating 
     stochastic one-step processes models is constructed. 
     Its application is illustrated by the predator-prey population dynamic system.

  \item[Conclusions] Computer algebra systems are very convenient for the purposes of
    rapid prototyping in mathematical models design and analysis.

  \end{description}

\end{abstract}


\maketitle

\section{Introduction}
\label{sec:intro}

This work corresponds our research on
mathematical models stochastization. This item is interesting due to
the following problems: the construction of population models from
first principles and the introduction of the stochastic into such models (the population dynamics 
is studied because of similar models introduction in other areas).

The problem of stochastic term introduction arises during mathematical models stochastization. 
There are several ways to solve this problem.
The easiest option is an 
in the deterministic equation. But when  additive stochastic term is introduced 
some free parameters that require further definition appears. Furthermore, these stochastic terms
usually interpreted as an external (rather than structural) random
impact. In this regard, we used and improved the
stochastic one--step processes models construction method, based on
master equation~\cite{L_lit10::en,L_lit14::en}. Stochastic differential
equation is considered as its approximate form. It allows
to get the model equations from general principles. Furthermore,
deterministic and stochastic parts are derived from the one 
equation so we can regard it as stochastic and deterministic parts consistency.
.

The aim of this work is the software complex development  for
rapid prototyping construction of stochastic
one--step processes models.  This complex consists of two blocks.
The first block generates the equations of
dynamic stochastic process model on the principles similar to chemical kinetic relations
describing the investigated process . This block is implemented
by means of the computer algebra system - system FriCAS, which is offshoot of Axiom.

The second block is used for the numerical analysis of the resulting model. For
numerical solution of deterministic and stochastic
models equations some Runge--Kutta different
orders methods~\cite{L_lit01,L_lit04} are used.

To illustrate the developed system the
well-known population model
predator--prey is used~\cite{p_5,p_6::en,p_25::en}.

The structure of the paper is as follows. The basic notation and conventions are introduced 
in Section \ref{sec:notation} 
Section~\ref{sec:onestep} is devoted to
brief introduction to the one--step processes stochastization method. Further,
in the Section~\ref{sec:pp-model} the model under investigation is described. In the
subsection~\ref{sec:pp-model:det} there is a brief reference to
standard (deterministic) approach, and in the
subsection~\ref{sec:pp-model:stoch} the stochastic
extension of our model with the help of the one--step
processes stochastization method is obtained.

In the Section~\ref{sec:axiom:compare} we justify selection of the system, which
implements the model equations generating unit. The actual interface
of this part of the program complex is described in the
Section~\ref{sec:axiom:realization}  .

 The possibility of applying
Runge--Kutta methods for the analysis of stochastic differential
equations is considered in the Section~\ref{sec:numerical}. The software interface of the model
equations numerical analysis unit is also described in this section. Calculations example is based on the 
predator--prey model.

\section{Notations and conventions}
\label{sec:notation}

\begin{enumerate}

\item We use abstract indices notation~\cite{penrose-rindler-1987::en}.
   In this notation tensor as 
   a whole object is denoted just as an index (e.g., $x^{i}$), 
   components are denoted by underlined index (e.g., 
   $x^{\crd{i}}$).

\item We will adhere to the following agreements. Latin indices of
  the middle of the alphabet ($i$, $j$, $k$) will apply to the space of
  the system state vectors. Latin indices from the beginning of the alphabet
  ($a$) will relate to the Wiener
  process space. Latin indices from the end of the alphabet ($p$, $q$) will
  refer to the indices of the Runge--Kutta method. Greek indices
  ($\alpha$) will set a number of different interactions in
  kinetic equations. 

\item A Dot over a symbol denotes differentiation with respect to time. 

\item The comma in the index denotes partial derivative with respect to 
   corresponding coordinate.
\end{enumerate}

\section{One-step processes modeling}
\label{sec:onestep}

Let's briefly review the method of one-step
processes stochastization on the basis of~\cite{kulyabov:2014:vestnik-miph:onestep::en}. 

We understand one-step processes as Markov processes with
continuous time with values in the domain of integers,
which transition matrix allows only transitions between neighbouring
portions. Also, these processes are known as birth-and-death processes.

One-step processes are subject to the following conditions:
\begin{enumerate}
\item If at the moment  $t$  the system is in state  $i \in
  \setZ_{\geqslant 0}$,then the probability of transition to state  
  $i+1$ in time interval $[t,t+\Delta t]$
 is equal to $k^{+} \Delta t + o (\Delta t)$.
\item If at time moment  $t$ the system is in state  $i \in
  \setZ_{+}$,then the probability of transition to state $i-1$ 
  in the time interval $[t,t+\Delta t]$ is equal to
  $k^{-} \Delta t + o (\Delta t)$.
\item The probability of transition to a state other than the neighbouring 
is equal to $o (\Delta t)$.
\item The probability to remain in the same state is equal to 
$1 - (k^{+} +  k^{-}) \Delta t + o (\Delta t)$.
\item State $ i = 0 $ is an absorbing boundary.
\end{enumerate}

The idea of the one-step processes stochastization method is as follows. 
Based on the patterns of interaction we
construct a master kinetic equation, expand it into a series,
leaving only the terms up to and including the second derivative. The resulting
equation is the Fokker--Planck equation. In order to get more convenient model we record 
corresponding Langevin equation. In fact, 
as we shall see , from the patterns of interaction we
will immediately obtain the coefficients of the Fokker--Planck equation (and
accordingly, the Langevin equation), so for practical use of the method 
there is no need to construct the master kinetic equation.

\subsection{Interaction schemes}

We will describe the state of the system by a state vector $x^{i} \ in
\setR^n $, where $n$ is the system dimension
(the state vector is considered as the set of mathematical values, fully describing system).
 The operator $ n^{i}_{j} \in \setZ^{n}_{{} \geqslant 0} \times \ setZ^{n}_{{} \geqslant 0} $ defines
the state of the system before the interaction and the operator $ m^{i}_{j} \in
\setZ^{n}_{{} \geqslant 0} \times \setZ^{n}_{{} \geqslant 0} $ 
--- after.the interaction The result of interaction is a system transition to 
another state ~ \cite {L_lit13, L_lit10::en}.

There are $s$ kinds of different interactions that may happen in the system, $s
\in \setZ_{+}$. So, instead of $n^{i}_{j}$ and $m^{i}_{j} $
let's consider the operators $ n^{i \alpha}_{j} \in
\setZ^{n}_{{}\geqslant 0} \times \setZ^{n}_{{}\geqslant 0} \times \setZ^{s}_{{}\geqslant 0} $ and
$m^{i \alpha}_{j} \in \setZ^{n}_{{}\geqslant 0} \times\setZ^{n}_{{}\geqslant 0} \times \setZ^{s}_{{}\geqslant 0} $.

System elements interaction will be described with the 
interaction schemes similar to chemical kinetic schemes~ \cite {lit_sdu_30::en}:
\begin{equation}
  \label{eq:chemkin}
  n^{i \alpha}_{j} x^j
  \overset{k_{\alpha}^{+}}{\underset{k_{\alpha}^{-}}{\rightleftharpoons}}
  m^{i \alpha}_{j} x^{j},
\end{equation}
Here Greek indexes specify the number of interactions and Latin ones specify
dimensionality of the system.
The state change is given by the operator
\begin{equation}
  \label{eq:r_i}
  r_j^{i \alpha} = m_j^{i \alpha} -n_j^{i \alpha}.
\end{equation}

Thus, one step interaction $ \crd {\alpha} $ in forward and
opposite directions can be written as
\begin{equation}
  \begin{gathered}
    x^{i}  \rightarrow x^i + r^{i \crd{\alpha}}_{j} x^{j},\\
    x^{i} \rightarrow x^{i} - r^{i \crd{\alpha}}_{j} x^{j}.
  \end{gathered}
\end{equation}

We can write~\eqref{eq:chemkin} not in the form of vector 
equations, but in the more traditional form sums:
\begin{equation}
  \label{eq:chemkin2}
  n^{i \alpha}_{j} x^j \delta_i
  \overset{k_{\alpha}^{+}}{\underset{k_{\alpha}^{-}}{\rightleftharpoons}}
  m^{i \alpha}_{j} x^{j} \delta_i,
\end{equation}
where $\delta_{\crd{i}} = (1,\ldots,1)$.

Also, we will use the following notation:
\begin{equation}
  \label{eq:n^i-notion}
  n^{i \alpha} := n^{i \alpha}_{j} \delta^{j}, 
  \quad m^{i \alpha} := m^{i \alpha}_{j} \delta^{j}, \quad
  r^{i \alpha} := r^{i \alpha}_{j} \delta^{j}.
\end{equation}

\subsection{Master equation}

Transition probabilities per unit of time from the state $x^{i}$ to
the state $x^{i} + r^{i \crd{\alpha}}_{j} x^{j}$ (to the state $x^{i} -
r^{i \crd{\alpha}}_{j} x^{j}$) are proportional to the number
of $x^{i}$ combination from a set of $n^{i \alpha}$ elements (of $x^{i}$ --- 
combinations from a set of $m^{i \alpha}$) and are given by:
\begin{equation}
  \label{eq:s-pm}
\begin{gathered}
  s^+_{\alpha} =  k^{+}_{\alpha} \prod_{\crd{i}=1}^{n}
  \frac{x^{\crd{i}}!}{(x^{\crd{i}} - n^{\crd{i} \alpha})!},       \\
  s^-_{\alpha} = k^{-}_{\alpha} \prod_{\crd{i}=1}^{n}
  \frac{x^{\crd{i}}!}{(x^{\crd{i}}-m^{\crd{i} \alpha})!}.
\end{gathered}
\end{equation}

Thus, the general form of the master kinetic equation for the states 
vector $x^{i}$ (it changes by $r^{i \crd{\alpha}}_j
x^j$ per step), takes the form:
\begin{multline} 
  \label{eq:uu}
  \frac{\partial p(x^{i} ,t)}{\partial t} 
  = {} \\ {} =
  \sum_{\crd{\alpha}=1}^{m} \left\{
    \left[s_{\crd{\alpha}}^{-}(x^{i}+r^{i \crd{\alpha}},t) p(x^{i}+r^{i \crd{\alpha}} ,t) -
      s_{\crd{\alpha}}^{+}(x^{i}) p(x^{i},t)
    \right] + \right. \\
  \left. + \left[ s_{\crd{\alpha}}^{+}(x^{i}-r^{i \crd{\alpha}}, t) p(x^{i} - r^{i
        \crd{\alpha}},t) - s_{\crd{\alpha}}^{-}(x^i) p(x^{i},t) \right] \right\}.
\end{multline}

\subsection{Fokker--Planck equation}

UWith the help of the Kramers--Moyal expansion,  the
Fokker--Planck equation~\cite{lit_sdu_30::en} is obtained. For this purpose we will make 
several assumptions:
\begin{enumerate}
\item there are only small jumps, ie  $s_{\alpha}(x^{i})$
  is a slowly varying function with the change of $x^{i}$;
\item $p(x^{i} ,t)$ also slowly changes with the change of $x^{i}$.
\end{enumerate}
Then in Fokker--Planck equation \eqref{eq:uu} one can shift  from the point $(x^{i} \pm r^{i
  \crd{\alpha}}_j x^j)$ to the point $x^{i}$, and by  expanding the right--hand 
side in a Taylor series and dropping terms of order higher than the second, we obtain
Fokker--Planck equation:
\begin{equation}
  \label{eq:FP}
  \frac{\partial p}{\partial t} = -
  \partial_{i} \left[ A^{i} p \right] + \frac{1}{2} \partial_{i} \partial_{j} \left[
    B^{i j}p \right],
\end{equation}
where
\begin{equation} 
  \label{eq:kFP}
  \begin{gathered}
    A^{i} := A^{i}(x^{k}, t) = r^{i \crd{\alpha}} \left[ s^+_{\crd{\alpha}} - s^-_{\crd{\alpha}} \right], \\
    B^{i j} := B^{i j}(x^{k},t)  = r^{i \crd{\alpha}} r^{j \crd{\alpha}} 
    \left[ s^+_{\crd{\alpha}} - s^-_{\crd{\alpha}} \right], \qquad
    \crd{\alpha} = \overline{1,m}.
  \end{gathered}
\end{equation}

As seen from \eqref{eq:kFP}, the coefficients of the Fokker--Planck equation
can be obtained directly from \eqref{eq:r_i} and \eqref{eq:s-pm}, i.e. in 
practical calculations, there is no need to write the master equation.

\subsection{Langevin equation}

The Langevin equation corresponds to the Fokker--Planck equation:
\begin{equation}
  \label{eq:langevin}
  \d x^{i} = a^{i} \d t + b^i_{a} \d W^{a},
\end{equation}
where $a^{i} := a^{i} (x^k, t)$, $b^{i}_{a} := b^{i}_{a} (x^k, t)$, $x^i
\in \setR^n $ --- is the system state vector, $W^{a} \in \mathbb{R}^m$
--- $m$-dimensional Wiener process. Wiener process is implemented as
$\d W = \varepsilon \sqrt{\d t}$, where $\varepsilon \sim
N(0,1)$~is normal distribution with average  $0$ and variance
$1$. Latin indexes from the middle of the alphabet denote the values related 
to the state vectors (dimension of the space is  $n$), and
Latin indexes from the beginning of the alphabet denote the values
related to the Wiener process vector (dimension of the space is
$m \leqslant n$).

The connection between the equation \eqref{eq:FP} and \eqref{eq:langevin}
expressed by the following relationships:
\begin{equation}
  \label{eq:k-langevin}
  A^{i} = a^{i}, \qquad B^{i j} = b^{i}_{a} b^{j a}.
\end{equation}

We will use Ito interpretation. Under the Ito interpretation, differential of complex functions does not
obey the standard formulas of analysis.
To calculate it \emph{rule} or \emph{Ito lemma} are used.

Let $f := f(x^{k},t)$ is a function of a random process $x^{k}(t)$,
$f\in\mathrm{C}^{2}$. Then the formula of the differential is~\cite{sdu_10}:
\begin{equation}
  \label{eq:lemma_ito}
  \d f = \left[\partial_{t}f + a^{i}f_{,i} +
    \frac{1}{2}b^{i}_{a}b^{i a}f_{,ij}\right]\d t +
  b_{a}^{i}f_{,i}\d W^{a},
\end{equation}
where $f:=f(x^{k},t)$, $a^{i}:=a^{i}(x^{k},t)$,
$b^{i}_{a}:=b^{i}_{a}(x^{k},t)$, and $\d W^{a} := \d
W^{a}(t)$.

\section{Predator--prey model}
\label{sec:pp-model}

\subsection{Deterministic predator--prey model}
\label{sec:pp-model:det}

Systems with the interaction of two predator-prey populations types are
extensively studied and there are a lot of various models for these systems.
The very first predator-prey model is considered to be a model which was obtained
independently by A.~Lotka and V.~Volterra. Lotka in~\cite{lotka} described some
hypothetical chemical reaction:
\begin{equation} 
\label{eq:0_4} 
A \xrightarrow{k_1}X \xrightarrow{k_2}
  Y \xrightarrow{k_3} B
\end{equation}
where $X,Y$ are intermediates substances, coefficients $k_1, k_2, k_3$ are
rates of chemical reactions, $A$ is a initial reagent, and $B$ is a
resultant. As a result was a system of differential
equations:
\begin{equation} 
  \label{eq:0_5} 
  \begin{cases}
    \Dot{x} = k_1 x - k_{2} x y, \\
    \Dot{y} = k_{2} x y - k_3 y.
  \end{cases}
\end{equation}

This system is identical to the system of differential equations, 
obtained by Volterra, who considered the growth mechanism of 
two populations with  predator--prey interaction type. In order to get equations~\cite{p_5}
Volterra made a series of idealized assumptions about 
nature of intraspecific and interspecific relationships in the  
predator--prey systenm 

\subsection{Stochastic predator--prey model}
\label{sec:pp-model:stoch}

Consider a model of predator--prey system, consisting of two individuals
species, one of which hunts, second is provided with
inexhaustible food resources. Let's introduce the notation, where $X$ is a prey
and $Y$ is a predator, then we can write the possible processes~\eqref{eq:chemkin2}
for the state vector $x^{\crd {i}} =
(X,Y)^T$~\cite{L_lit09::en,L_lit11::en,kulyabov:2013:conference:mephi::en,L_lit12::en}:
\begin{equation} 
  \label{eq:3_0}
  \begin{aligned}
    X &\ \xrightarrow{k_1} 2X, &\ r^{\crd{i} 1}=(1,0)^T \\
    X + Y &\ \xrightarrow{k_2} 2Y, &\ r^{\crd{i} 2} = (-1,1)^T \\
    Y &\ \xrightarrow{k_3} 0, &\ r^{\crd{i} 3}=(0,-1)^T,
  \end{aligned}
\end{equation}
which have the following interpretation. The first relation means
that the prey which eats the food unit immediately
reproduced. The second relation describes the case when  predator absorbs the prey
and than it is instantaneous reproduced. Only such possibility of prey death is considered.
Last ratio is a natural predator death.

All processes are irreversible, so $s^{-}_{\alpha} = 0$, and
\begin{equation}
  \begin{gathered}
    s^+_1(x,y)= k_1 \frac{x!}{(x-1)!}\frac{y!}{y!}=k_1 x, \\
    s^+_2(x,y)= k_2\frac{x!}{(x-1)!}\frac{y!}{(y-1)!}=k_2 x y, \\
    s^+_3(x,y)= k_3\frac{x!}{x!}\frac{y!}{(y-1)!}=k_3 y.
  \end{gathered}
\end{equation} 

With the help of the formula~\eqref{eq:FP} we have the Fokker--Planck equation:
\begin{equation} 
  \label{eq:FP_hzh} 
  \frac{\partial p(x,y)}{\partial t} = 
  -\partial_i \left( A^{i} (x,y) p(x,y) \right) +
  \frac{1}{2} \partial_i\partial_j\left(B^{ij}(x,y) p(x,y)\right),
\end{equation}
where
\begin{equation}
  \begin{gathered}
    A^i(x,y) = s_{\alpha}^+ (x,y)\ r^{i \alpha},\\
    B^{ij}(x,y) = s_{\alpha}^+ (x,y) r^{i \alpha} r^{j \alpha}.
  \end{gathered}
\end{equation}

As a result:
\begin{equation}
\label{eq:pp_FP}
  \begin{multlined}
    A^{\crd{i}}(x,y)=
    \begin{pmatrix}
      1\\
      0
    \end{pmatrix}
    k_1 x +
    \begin{pmatrix}
      -1\\
      1
    \end{pmatrix}
    k_2xy + {} \\
                {} + 
    \begin{pmatrix}
      0\\
      -1
    \end{pmatrix}
    k_3y =
    \begin{pmatrix}
      k_1 x-k_2xy\\
      k_2xy - k_3y
    \end{pmatrix}, \\
      B^{\crd{i} \crd{j}}(x,y) =
      \begin{pmatrix}
        1\\
        0
      \end{pmatrix}
      (1,0) k_1 x +
      \begin{pmatrix}
        -1\\
        1
      \end{pmatrix}
      (-1,1) k_2xy + {} \\
                        {} + 
      \begin{pmatrix}
        0\\
        -1
      \end{pmatrix}
      (0,-1) k_3y  =
      \begin{pmatrix}
        k_1 x+k_2xy & -k_2xy \\
        -k_2xy & k_2xy + k_3y
      \end{pmatrix}.
  \end{multlined}
\end{equation}

In order to write a stochastic differential equation in 
Langevin form~\eqref{eq:langevin} for predator--prey model, it is
enough to take the square root of the resulting matrix $B^{ij}$ in 
Fokker--Planck equation.
\begin{equation}
  \label{eq:pp_langevin}
  \begin{gathered}
  \d 
  \begin{pmatrix}
    x \\
    y
  \end{pmatrix}
  =
  \begin{pmatrix}
    k_1 x-k_2xy\\
    k_2xy - k_3y
  \end{pmatrix}\d t + 
        b^{\crd{i}}_{\crd{a}} 
  \begin{pmatrix}
    \d W^{1} \\
    \d W^{2}
  \end{pmatrix},\\
  b^{\crd{i}}_{a} b^{\crd{j} a} = B^{\crd{i}\crd{j}} =
    \begin{pmatrix}
      k_1 x+k_2xy & -k_2xy \\
      -k_2xy & k_2xy + k_3y
    \end{pmatrix}.
        \end{gathered}
\end{equation}

It should be noted that the specific form of the matrix $b^{i}_{a}$ is not written out
because of the extreme awkwardness of the expression. However, with further 
studies we will need not actually matrix $b^{i}_{a}$, but its 
square, i.e. the matrix $B^{ij}$.

\section{Implementation of the one-step stochastic processes model in the computer algebra system}
\label{sec:axiom}

\subsection{Justification of the computer algebra system choice}
\label{sec:axiom:compare}

Let's consider systems of analytical calculations, Maxima and Axiom.
Maxima is the first system of analytical calculations
and it is written in Lisp. Maxima successfully runs on all modern
operating systems: Windows, Linux and UNIX, Mac OS and even on PDA
running Windows CE/Mobile. Documentation is integrated into the program as
a handbook with search. There is no distinction between
objects and data in Maxima, and there is no clear distinction between the operator and function.
There is no integrated graphics rendering in the system.

Unlike Maxima Axiom language is strongly typified for better mathematical objects and 
relationships display. The mathematical basis is written in
Spad language . Axiom portability is slightly worse: the system runs under Linux,
UNIX, and graphs does not work under Windows. Axiom has its own
graphics subsystem.

In 2007 two Axiom open source forks appeared: OpenAxiom and
FriCAS. Open Axiom is developed by adhering to the ideology of Axiom, problems that occurred 
in the Axiom are eliminated.
FriCAS developers reorganized the assembly process, expanded functionality.
Furthermore, FriCAS supports not only GCL, which operates on limited
number of platforms, but ECL, Clisp, sbcl or openmcl,
that allows to run FriCAS  under wider range of platforms.

\subsection{Implementation description in the Axiom computer algebra system}
\label{sec:axiom:realization}

\begin{figure}
   \centering
   \includegraphics[width=0.9\linewidth]{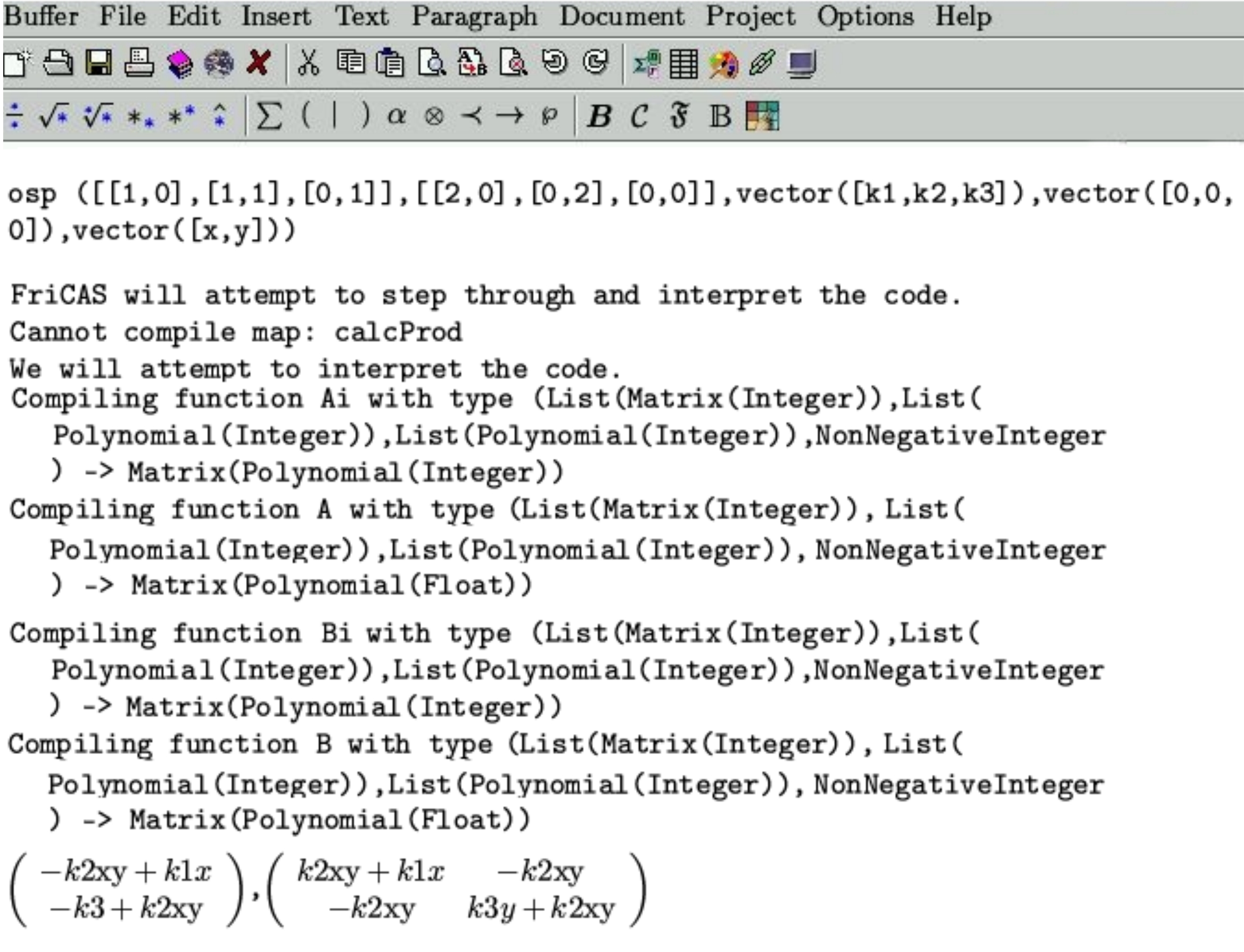}
   \caption{The output of the module for predator--prey model in
     graphic \TeX{}macs shell}
\label{fig:texmacs:pp_FP}
 \end{figure}

Method of one-step processes randomization is organized as a module
for the FriCAS computer algebra system.
To display all the calculations on the screen the variable \verb|SHOWCALC:=true| is used.
To call the method you need to use the main function, which has the following
view:
\begin{verbatim}
osp(Matrix(Integer), Matrix(Integer), 
        Vector, Vector, Vector)
\end{verbatim}
where the first argument is before interaction states matrix $n^{i}_{j}$, 
the second argument is after interaction states matrix $m^{i}_{j}$, 
the third argument is the vector $k^{+}_{\alpha}$, 
the fourth argument is the vector $k^{-}_{\alpha}$, 
the fifth argument is the state vector $x^{i}$.
Let's consider the features of the language FriCAS on auxiliary functions.
For example, the function \verb|calcProd| is used to simplify the calculations
$s^{+}_{\alpha}$ and $s^{-}_{\alpha}$. In the implementation of the function
operator of the condition and built-in function reduce are used:
\begin{verbatim}
calcProd : (Matrix(Integer), Vector, Integer, 
        Integer) -> Void
calcProd (n, x, a, i) == 
        nai:Integer := n(a,i) 
        if nai = 0 then 1 else reduce(*,
                [x(i) - j for j in 0..(nai-1)])
\end{verbatim}

In the function \verb|Bi| intermediate calculations for
elements of the matrix $B^{ij}$ are made:
\begin{verbatim}
Bi (rv, sp, sm, i) == rv(i) * 
        (transpose rv(i)) * (sp(i) + sm(i))
\end{verbatim}

In order to use the module for predator--prey system model, we
call the function with the following arguments:
\begin{verbatim}
osp ([[1,0],[1,1],[0,1]],[[2,0],[0,2],[0,0]],
        vector([k1,k2,k3]), vector([0,0,0]),
               vector([x,y]))
\end{verbatim}

Fig.~\ref{fig:texmacs:pp_FP} represents the result obtained in \TeX{}macs
shell. In fact, we repeated the results obtained in~\eqref{eq:pp_FP}.

\section{Numerical experiment for the program complex}
\label{sec:numerical}

\subsection{Stochastic Runge--Kutta methods}

Euler--Maruyama method is one of well-known numerical methods for solving SDE,
it is a special case of a more general
Stochastic Runge--Kutta method. Classical Runge--Kutta method
can be generalized to the case of the SDE system~\eqref{eq:langevin} in the following 
manner~\cite{L_lit01, L_lit04}:
\begin{equation}
  \left\{
    \begin{aligned}
      X^{i}_{k} = & x_{0}^{i} +
      hR^{\crd{l}}_{k}a^{i}(X^{1}_{\crd{l}},\ldots,X^{n}_{\crd{l}}) +
      \widehat{R}^{\crd{l}}_{k}J^{\alpha}b^{i}_{\alpha}(X^{1}_{\crd{l}},\ldots,X^{n}_{\crd{l}}),\\
      x^{i}_{1} = & x_{0}^{i} +
      hr^{\crd{l}}a^{i}(X^{1}_{\crd{l}},\ldots,X^{n}_{\crd{l}}) +
      \hat{r}^{\crd{l}}J^{\alpha}b^{i}_{\alpha}(X^{1}_{\crd{l}},\ldots,X^{n}_{\crd{l}})
    \end{aligned}
  \right.
\end{equation}

Indexes $k = 1,\ldots,s$ and $l = 1,\ldots,n$ refer to
stochastic Runge--Kutta method. $J \sim N(0,h)$ or $J \sim
\sqrt{h}\varepsilon$, $\varepsilon \sim N(0,1)$ are normal
distributed random variables. Such a choice of these numerical values for
approximation is made because the Wiener process is implemented as
$\d W = \varepsilon \sqrt{\d t}$. You should also pay attention to
double summation in the third term of both numerical scheme formulas
as well as the fact that each number $J^{1},\ldots,J^{n}$ should
generated separately.

The method coefficients, as well as for the classical analogue, can be
grouped into a table called the \emph{Butcher table}:

\begin{equation}
\setstretch{1.8}
\begin{array}{c|c|c}
  &R_{ij}&        \hat{R}_{ij}\\ \hline
  &r_{j}&         \hat{r}_{j}
\end{array}.
\end{equation}

For calculations we used a method with the table

\begin{equation}
\setstretch{1.2}
\begin{array}{c|ccc|ccc}
  &0&     0&      0&      0&      0&      0\\
  &2/3&   0&      0&      2/3&    0&      0\\
  &-1&    1&      0&      -1&     1&      0\\ \hline
  & 0&    3/4&    1/4&    0&      3/4&    1/4
\end{array}
\end{equation}

\subsection{Software implementation description} 

The purposes of the programs complex were to  automate the SDE coefficients
$A^{i}$ and $B^{ij}$ computation with the help of general principles described above,
and to find a numerical solution of the equation obtained by means of stochastic
Runge--Kutta methods. From a programming standpoint we can derive 
three subtasks:
\begin{enumerate}
\item coefficients $A^{i}$ and $B^{ij}$ generation using the
  computer algebra system;
\item generation of source code in languages
  Fortran and Julia, implementing the SDE on the basis of the coefficients,
  saved as a text file;
\item writing subroutines/functions implementing stochastic
  Runge--Kutta methods in Fortran and Julia, and their subsequent compilation
  together with automatically generated source codes.
\end{enumerate}

As a result of its work Axiom module  creates a text file
which contains the coefficients $A^{i}$ and $B^{ij}$ in the following form:
\begin{verbatim}
        # A
        A[1]
        ...
        A[N]
        # B
        B[1,1] B[1,2] .. B[1,N]
        ...
        B[N,1] B[N,2] .. B[N,N] 
\end{verbatim} 
Matrix $b^{i}_{\alpha} = \sqrt{b^{i}_{\alpha}b^{j\alpha}} =
\sqrt{B^{ij}}$ is calculated numerically with the help of the singular value matrix decomposition 
(a subroutine DGESVD from library LAPACK is used).

For the second subtask scripting language Python was chosen 
(version 3). This language has a wide set of tools to work
with strings and text files. Except matrices $A^{i}$ and $B^{ij}$
additional information about the mathematical model was specified as
dictionary (standard data type in Python), with model name, list of variables, 
list of parameters, initial values of
variables, parameters values and parameters of the numerical method 
(integration section and step size).

On the basis of these data, the script automatically generates two files
\verb|functions.f90| and \verb|main.f90|, where the first is a module
with functions defining the SDE, and the second one is a main
program file. While compiling these files the third additional module with
auxiliary procedures with Stochastic Runge--Kutta method is added.

\subsection{The numerical experiment description}

For the programs complex work verification a well-known predator--prey model 
was chosen with vector $a^{i}$  components
\begin{equation}
  a^{1} = \alpha x - \beta xy,\;\; a^{2} = -\gamma y + \delta xy
\end{equation}
and matix $B^{ij}$:
\begin{equation}
  \left[
    \begin{array}{cc}
      \alpha x + \beta xy & -\beta xy \\
      -\beta xy & \beta xy + \gamma x
    \end{array}
  \right],
\end{equation}
$x$  is the number of preys, $ y $ is the number of predators. Coefficients 
also have the following physical (biological) meaning:
$\alpha$ is the growth rate of the prey population,
$\beta$ is a frequency of predators and prey meetings,
$\gamma$ is an intensity of predators death or migration in a lack of preys,
$\delta$ is a predator population growth rate on the assumption of the excess of the prey.

During numerical simulations it was taken into account that the value of variables $x,y$ could not 
be less than zero (program stop working when one of the variables becomes equal to zero).

\begin{figure}
  \centering
  \includegraphics[width=0.7\linewidth]{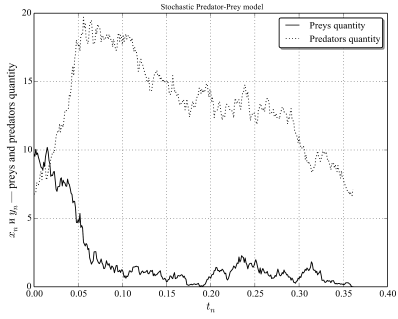}
\caption{Stochastic predator--prey model, prey die}
\label{fig:prepre_1_2}
\end{figure}

Numerical simulation shows that the addition of stochastic to the classical 
predator--prey model leads to the fact that after a certain time death of one of the competing species
comes. So, for the following parameters: 
$\alpha= 10$, $\beta= 1,5$, $\gamma= 8,5$, $\delta= 1,8$ and the initial 
values: $x = 9,7$, $x = 6,77$, victims are first to die, and after that 
predators die due to  lack. This case is illustrated in Fig.~\ref{fig:prepre_1_2}. 
For comparison in Fig.~\ref{fig:prepre_1_2d} is a graph for the deterministic case .

\begin{figure}
  \centering
  \includegraphics[width=0.7\linewidth]{img_1_2d}
\caption{Deterministic predator--prey model}
\label{fig:prepre_1_2d}
\end{figure}

\begin{figure}
   \centering
   \includegraphics[width=0.7\linewidth]{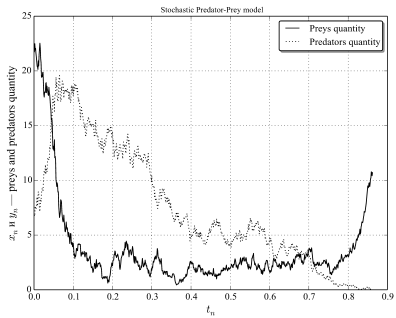}
\caption{Stochastic predator--prey model, predators die}
\label{fig:prepre_2_2}
\end{figure}

Under other conditions ($\alpha = 10$, $\beta = 1,5$, 
$\gamma = 8,5$, $\delta = 0,5$, $x = 22$, $y = 6,76$) 
predators die, and the number of victims is increasing rapidly, as for 
their model assumes an infinite source of food. Graphics for 
this case are shown in Fig.~\ref{fig:prepre_2_2}, and 
Fig.~\ref{fig:prepre_2_2d} shown for comparison with deterministic case.

\begin{figure}
  \centering
  \includegraphics[width=0.7\linewidth]{img_2_2d}
\caption{Deterministic predator--prey model}
\label{fig:prepre_2_2d}
\end{figure}

\section{Conclusions}
\label{sec:conclusion}

This work demonstrates the application of the developed initial physical system formalization method.
The system is presented in the form of one or more
one-step processes. Formalization of the system is done by introducing the evolution operator.
Wherein the analytical description of the model requires a lot of
routine operations. To simplify the work we propose to use the computer
algebra system (Axiom fork FriCAS).

We have developed an analytical software package block that receives
inlet evolution operator and produces the SDE, which describes
the original model. For numerical studies of obtained
SDE system a second software unit that converts the resulting system of equations 
into the program code in fortran and gives
its numerical solution was developed. Thus, the software system is applicable
for both analytical and numerical study of the original model.

Currently the software package does not cover all possibilities,
incorporated in the proposed method of formalizing the original physical system.
Since the original system description uses ODE, we should introduce the boundary conditions 
by ties or indicator functions.
Partial differential equations can help to solve this problem. 
Further objective is the development of a complete 
software complex for a method of one-step original physical system model construction.

 \bibliographystyle{unsrtnat}

\bibliography{bib/axiom-sdu/axiom-sdu}

\end{document}


\graphicspath{{image/axiom-sdu/ru/}}

\title{Программный комплекс стохастического моделирования одношаговых процессов}

\author{Е. Г. Еферина}
\email{eg.eferina@gmail.com}
\affiliation{Кафедра прикладной информатики и теории вероятностей,\\
  Российский университет дружбы народов,\\
  ул. Миклухо-Маклая, д.6, Москва, Россия, 117198}

\author{А. В. Королькова}
\email{avkorolkova@gmail.com} 
\affiliation{Кафедра прикладной информатики и теории вероятностей,\\
  Российский университет дружбы народов,\\
  ул. Миклухо-Маклая, д.6, Москва, Россия, 117198}

\author{М. Н. Геворкян}
\email{mngevorkyan@sci.pfu.edu.ru} 
\affiliation{Кафедра прикладной информатики и теории вероятностей,\\
  Российский университет дружбы народов,\\
  ул. Миклухо-Маклая, д.6, Москва, Россия, 117198}

\author{Д. С. Кулябов}
\email{yamadharma@gmail.com}
\affiliation{Кафедра прикладной информатики и теории вероятностей,\\
  Российский университет дружбы народов,\\
  ул. Миклухо-Маклая, д.6, Москва, Россия, 117198}
\affiliation{Лаборатория информационных технологий,\\
Объединённый институт ядерных исследований,\\
ул. Жолио-Кюри 6, Дубна, Московская область, Россия, 141980}

\author{Л. А. Севастьянов}
\email{leonid.sevast@gmail.com}
\affiliation{Кафедра прикладной информатики и теории вероятностей,\\
  Российский университет дружбы народов,\\
  ул. Миклухо-Маклая, д.6, Москва, Россия, 117198}
\affiliation{Лаборатория теоретической физики,\\
  Объединённый институт ядерных исследований,\\
  ул. Жолио-Кюри 6, Дубна, Московская область, Россия, 141980}

\thanks{Опубликовано в:
  One-Step Stochastic Processes Simulation Software Package~/
  E.~G. Eferina, A.~V. Korolkova, M.~N. Gevorkyan, D.~S. Kulyabov, L.~A.
  Sevastyanov~// Bulletin of Peoples’
  Friendship University of Russia. Series «Mathematics. Information Sciences.
  Physics».~--- 2014.~--- No.~3.~--- P.~46--59.
}

\thanks{Исходные тексты:
  \url{https://bitbucket.org/yamadharma/articles-2014-axiom-sdu}}

\begin{abstract}

  \begin{description}

  \item[Предпосылки] Нашим коллективом разработана методика
    согласованного (зависящего от
    структуры системы) введения стохастики в детерминистические
    модели. На данном этапе методика ограничена классом одношаговых
    процессов.

  \item[Цель] Для оптимизации работы по созданию моделей следует
    автоматизировать как можно больше рутинных операций. В данном
    случае процесс составления уравнений модели можно
    алгоритмизировать и реализовать в системе компьютерной
    алгебры. Кроме того, на базе этих результатов можно получить и
    набор программ для проведения численного эксперимента.

  \item[Методы] Для реализации аналитических расчётов используется
    система компьютерной алгебры Axiom. Для проведения численного
    эксперимента используются языка FORTRAN и Julia. В качестве
    численного метода используется метод Рунге--Кутты для
    стохастических дифференциальных уравнений.

  \item[Результаты] Создан программный комплекс для создания
    стохастических моделей одношаговых процессов. Проиллюстрировано
    его применение на примере системы популяционной динамики типа
    <<хищник--жертва>>. Детерминистические модели для таких процессов
    достаточно хорошо исследованы, что позволяет сравнить полученные
    результаты с уже известными.

  \item[Выводы] Системы компьютерной алгебры очень удобны для целей
    быстрого прототипирования при создании и исследовании
    математических моделей.

  \end{description}

\end{abstract}

  \keywords{стохастические дифференциальные уравнения; модель
    <<хищник--жертва>>; основное кинетическое уравнения; уравнение
    Фоккера--Планка;  системы компьютерной алгебры; система Axiom}

\maketitle

\section{Введение}
\label{sec:intro}

Данная работа находится в русле проводимых нами исследований по
стохастизации математических моделей. Данная тематика интересна из-за 
следующих проблем: построение популяционных моделей из
первых принципов и введение стохастики в модели данного вида (следует
заметить, что обратились мы к популяционной динамике потому, что нами
исследовались схожие модели в других областях).

При стохастизации математических моделей возникает проблема введения
стохастического члена. Существует несколько способов сделать это.
Самый простой вариант --- аддитивное добавление стохастического члена
к детерминистическому уравнению. Однако при таком введении возникают
свободные параметры, требующие дальнейшего определения из
дополнительных соображений. Кроме того, данные стохастические члены
обычно интерпретируются как внешнее (а не структурное) случайное
воздействие.  В связи с этим мы использовали и усовершенствовали метод
построения стохастических моделей одношаговых процессов на основе
основного кинетического
уравнения~\cite{L_lit10,L_lit14}. Стохастическое дифференциальное
уравнение рассматривается как его приближённая форма. Это позволяет
получить модельные уравнения из общих принципов. Кроме того,
детерминистическая и стохастическая части получаются из одного и того
же уравнения, что рассматривается нами как согласованность
стохастической и детерминистической частей.

Целью данной работы является разработка программного комплекса для
быстрого прототипирования построения стохастических моделей
одношаговых процессов. Данный комплекс состоит из двух блоков. 
Первый блок по соотношениям, аналогичным соотношениям химической 
кинетики, описывающим исследуемый процесс, генерирует уравнения 
динамической стохастической модели процесса. Этот блок реализован 
на системе компьютерной алгебры. В качестве системы для реализации 
выбрана система FriCAS~---ответвление системы Axiom.

Второй блок служит для численного анализа полученной модели. Для
численного решения уравнений детерминированной и стохастической
моделей используются методы Рунге--Кутты разных
порядков~\cite{L_lit01, L_lit04}. 

Для демонстрации работы разработанного комплекса используется 
хорошо известная популяционная модель 
<<хищник--жертва>>~\cite{p_5, p_6, p_25}.

Структура статьи следующая. В разделе \ref{sec:notation} введены
основные обозначения и соглашения. В разделе~\ref{sec:onestep} даётся
краткое введение в метод стохастизации одношаговых процессов. Далее, в
разделе~\ref{sec:pp-model} описывается исследуемая модель. При этом в
подразделе~\ref{sec:pp-model:det} даётся краткая справка по
стандартному (детерминистическому) подходу, а в
подразделе~\ref{sec:pp-model:stoch} мы получаем стохастическое
расширение данной модели по методу стохастизации одношаговых
процессов.

В разделе~\ref{sec:axiom:compare} обосновывается выбор системы для
реализации блока генерации уравнений модели. В
разделе~\ref{sec:axiom:realization} описывается собственно интерфейс
работы с этой частью программного комплекса.

В разделе~\ref{sec:numerical} рассматривается возможность применения
методов Рунге--Кутты для анализа стохастических дифференциальных
уравнений и описывается программный интерфейс блока численного анализа
модельных уравнений. Пример расчётов основан на модели
<<хищник--жертва>>.

\section{Обозначения и соглашения}
\label{sec:notation}

\begin{enumerate}

\item В работе используется нотация абстрактных
  индексов~\cite{penrose-rindler-1987}. В данной нотации тензор как
  целостный объект обозначается просто индексом (например, $x^{i}$),
  компоненты обозначаются подчёркнутым индексом (например,
  $x^{\crd{i}}$).

\item Будем придерживаться следующих соглашений. Латинские индексы из
  середины алфавита ($i$, $j$, $k$) будут относиться к пространству
  векторов состояний системы. Латинские индексы из начала алфавита
  ($a$) будут относиться к пространству винеровского
  процесса. Латинские индексы из конца алфавита ($p$, $q$) будут
  относиться к индексам метода Рунге--Кутты. Греческие индексы
  ($\alpha$) будут задавать количество разных взаимодействий в
  кинетических уравнениях.

\item Точкой над символом обозначается дифференцирование по времени.

\item Запятой в индексе обозначается частная производная по
  соответствующей координате.
\end{enumerate}

\section{Моделирование одношаговых процессов}
\label{sec:onestep}

Дадим краткий обзор метода стохастизации одношаговых процессов на базе
работы~\cite{kulyabov:2014:vestnik-miph:onestep}.

Под одношаговыми процессами мы будем понимать марковские процессы с
непрерывным временем, принимающие значения в области целых чисел,
матрица перехода которых допускает только переходы между соседними
участками. Также эти процессы известны под названиями процессов
рождения--гибели. 

Одношаговые процессы подчиняются следующим условиям.
\begin{enumerate}
\item Если в момент времени $t$ система находится в состоянии $i \in
  \setZ_{\geqslant 0}$, то вероятность перехода в состояние $i+1$ в интервале
  времени $[t,t+\Delta t]$ равна $k^{+} \Delta t + o (\Delta t)$.
\item Если в момент времени $t$ система находится в состоянии $i \in
  \setZ_{+}$, то вероятность перехода в состояние $i-1$ в интервале
  времени $[t,t+\Delta t]$ равна $k^{-} \Delta t + o (\Delta t)$.
\item Вероятность перехода в состояние, отличное от соседних равна $o
  (\Delta t)$.
\item Вероятность сохранения прежнего состояния равна $1 - (k^{+} +
  k^{-}) \Delta t + o (\Delta t)$.
\item Состояние $i = 0$ есть поглощающая граница.
\end{enumerate}

Идея метода стохастизаци одношаговых процессов состоит в следующем. 
На основании схем взаимодействия мы
строим основное кинетическое уравнение, раскладываем его в ряд,
оставляя только члены до второй производной включительно. Получившееся
уравнение будет уравнением Фоккера--Планка. Для получения более
привычного вида модели записываем соответствующее ему уравнение
Ланжевена. На самом деле, как мы увидим, из схем взаимодействия мы
сразу получаем коэффициенты уравнения Фоккера--Планка (и,
соответственно, уравнения Ланжевена), поэтому при практическом
применении метода строить основное кинетическое уравнение нет
необходимости. 

\subsection{Схемы взаимодействия}

Состояние системы будем описывать вектором состояния $x^{i} \in
\setR^n$, где $n$~--- размерность системы (под вектором состояния
будем понимать множество математических величин, полностью описывающих
систему). Оператор $n^{i}_{j} \in \setZ^{n}_{{}\geqslant 0} \times \setZ^{n}_{{}\geqslant 0}$ задаёт
состояние системы до взаимодействия, оператор $m^{i}_{j} \in
\setZ^{n}_{{}\geqslant 0} \times \setZ^{n}_{{}\geqslant 0}$~--- после. В результате
взаимодействия происходит переход системы в другое
состояние~\cite{L_lit13, L_lit10}.

В системе может происходить $s$ видов различных взаимодействий, где $s
\in \setZ_{+}$. Поэтому вместо операторов $n^{i}_{j}$ и $m^{i}_{j}$
будем рассматривать операторы $n^{i \alpha}_{j} \in
\setZ^{n}_{{}\geqslant 0} \times \setZ^{n}_{{}\geqslant 0} \times \setZ^{s}_{{}\geqslant 0}$ и
$m^{i \alpha}_{j} \in \setZ^{n}_{{}\geqslant 0} \times \setZ^{n}_{{}\geqslant 0} \times \setZ^{s}_{{}\geqslant 0}$.

Взаимодействие элементов системы будем описывать с помощью схем
взаимодействия, подобным схемам химической кинетики~\cite{lit_sdu_30}:
\begin{equation}
  \label{eq:chemkin}
  n^{i \alpha}_{j} x^j
  \overset{k_{\alpha}^{+}}{\underset{k_{\alpha}^{-}}{\rightleftharpoons}}
  m^{i \alpha}_{j} x^{j},
\end{equation}
Здесь греческие индексы задают количество взаимодействий, а латинские~---
размерность системы. 
Изменение состояния будет задаваться оператором
\begin{equation}
  \label{eq:r_i}
  r_j^{i \alpha} = m_j^{i \alpha} -n_j^{i \alpha}.
\end{equation}

Таким образом, один шаг взаимодействия $\crd{\alpha}$ в прямом и
обратном направлениях можно записать в виде
\begin{equation}
  \begin{gathered}
    x^{i}  \rightarrow x^i + r^{i \crd{\alpha}}_{j} x^{j},\\
    x^{i} \rightarrow x^{i} - r^{i \crd{\alpha}}_{j} x^{j}.
  \end{gathered}
\end{equation}

Мы также можем записывать~\eqref{eq:chemkin} не в форме векторных уравнений, 
а в виде более традиционных сумм:
\begin{equation}
  \label{eq:chemkin2}
  n^{i \alpha}_{j} x^j \delta_i
  \overset{k_{\alpha}^{+}}{\underset{k_{\alpha}^{-}}{\rightleftharpoons}}
  m^{i \alpha}_{j} x^{j} \delta_i,
\end{equation}
где $\delta_{\crd{i}} = (1,\ldots,1)$.

Также мы будем использовать следующие обозначения:
\begin{equation}
  \label{eq:n^i-notion}
  n^{i \alpha} := n^{i \alpha}_{j} \delta^{j}, 
  \quad m^{i \alpha} := m^{i \alpha}_{j} \delta^{j}, \quad
  r^{i \alpha} := r^{i \alpha}_{j} \delta^{j}.
\end{equation}

\subsection{Основное кинетическое уравнение}

Вероятности перехода в единицу времени из состояния $x^{i}$ в
состояние $x^{i} + r^{i \crd{\alpha}}_{j} x^{j}$ (в состояние $x^{i} -
r^{i \crd{\alpha}}_{j} x^{j}$) пропорциональны соответственно числу
способов выбора комбинации из $x^{i}$ по $n^{i \alpha}$ (из $x^{i}$ по
$m^{i \alpha}$) и определяются выражениями:
\begin{equation}
  \label{eq:s-pm}
\begin{gathered}
  s^+_{\alpha} =  k^{+}_{\alpha} \prod_{\crd{i}=1}^{n}
  \frac{x^{\crd{i}}!}{(x^{\crd{i}} - n^{\crd{i} \alpha})!},       \\
  s^-_{\alpha} = k^{-}_{\alpha} \prod_{\crd{i}=1}^{n}
  \frac{x^{\crd{i}}!}{(x^{\crd{i}}-m^{\crd{i} \alpha})!}.
\end{gathered}
\end{equation}

Таким образом, общий вид основного кинетического уравнения для вектора
состояний $x^{i}$, изменяющегося шагами длины $r^{i \crd{\alpha}}_j
x^j$, принимает вид:
\begin{multline} 
  \label{eq:uu}
  \frac{\partial p(x^{i} ,t)}{\partial t} 
  = {} \\ {} =
  \sum_{\crd{\alpha}=1}^{m} \left\{
    \left[s_{\crd{\alpha}}^{-}(x^{i}+r^{i \crd{\alpha}},t) p(x^{i}+r^{i \crd{\alpha}} ,t) -
      s_{\crd{\alpha}}^{+}(x^{i}) p(x^{i},t)
    \right] + \right. \\
  \left. + \left[ s_{\crd{\alpha}}^{+}(x^{i}-r^{i \crd{\alpha}}, t) p(x^{i} - r^{i
        \crd{\alpha}},t) - s_{\crd{\alpha}}^{-}(x^i) p(x^{i},t) \right] \right\}.
\end{multline}

\subsection{Уравнение Фоккера--Планка}

Далее, используя разложение Крамерса--Мойала, получаем уравнение
Фоккера--Планка~\cite{lit_sdu_30}. Для этого делается
несколько предположений:
\begin{enumerate}
\item имеют место только малые скачки, т.е. $s_{\alpha}(x^{i})$
  является функцией, медленно изменяющейся с изменением $x^{i}$;
\item $p(x^{i} ,t)$ также медленно изменяется с изменением $x^{i}$.
\end{enumerate}
Тогда можно выполнить сдвиг в~\eqref{eq:uu} из точки $(x^{i} \pm r^{i
  \crd{\alpha}}_j x^j)$ в точку $x^{i}$ и, разложив правую часть в ряд
Тейлора и отбросив члены порядка выше второго, получим уравнение
Фоккера-Планка:
\begin{equation}
  \label{eq:FP}
  \frac{\partial p}{\partial t} = -
  \partial_{i} \left[ A^{i} p \right] + \frac{1}{2} \partial_{i} \partial_{j} \left[
    B^{i j}p \right],
\end{equation}
где
\begin{equation} 
  \label{eq:kFP}
  \begin{gathered}
    A^{i} := A^{i}(x^{k}, t) = r^{i \crd{\alpha}} \left[ s^+_{\crd{\alpha}} - s^-_{\crd{\alpha}} \right], \\
    B^{i j} := B^{i j}(x^{k},t)  = r^{i \crd{\alpha}} r^{j \crd{\alpha}} 
    \left[ s^+_{\crd{\alpha}} - s^-_{\crd{\alpha}} \right], \qquad
    \crd{\alpha} = \overline{1,m}.
  \end{gathered}
\end{equation}

Как видно из \eqref{eq:kFP}, коэффициенты уравнения Фоккера--Планка
можно получить сразу из \eqref{eq:r_i} и \eqref{eq:s-pm}, то есть в
практических расчётах записывать основное кинетическое уравнение нет
необходимости.

\subsection{Уравнение Ланжевена}

Уравнению Фоккера--Планка соответствует уравнение Ланжевена:
\begin{equation}
  \label{eq:langevin}
  \d x^{i} = a^{i} \d t + b^i_{a} \d W^{a},
\end{equation}
где $a^{i} := a^{i} (x^k, t)$, $b^{i}_{a} := b^{i}_{a} (x^k, t)$, $x^i
\in \setR^n $ --- вектор состояния системы, $W^{a} \in \mathbb{R}^m$
--- $m$-мерный винеровский процесс. Винеровский процесс реализуется
как $\d W = \varepsilon \sqrt{\d t}$, где $\varepsilon \sim
N(0,1)$~--- нормальное распределение со средним $0$ и дисперсией
$1$. Латинскими индексами из середины алфавита обозначаются величины,
относящиеся к векторам состояний (размерность пространства $n$), а
латинскими индексами из начала алфавита обозначаются величины,
относящиеся к вектору винеровского процесса (размерность пространства
$m \leqslant n$).

При этом связь между уравнениями \eqref{eq:FP} и \eqref{eq:langevin}
выражается следующими соотношениями:
\begin{equation}
  \label{eq:k-langevin}
  A^{i} = a^{i}, \qquad B^{i j} = b^{i}_{a} b^{j a}.
\end{equation}

Мы будем использовать интерпретацию Ито.
В рамках интерпретации Ито дифференциал от сложной функции не
подчиняется стандартным формулам анализа. Для его вычисления
используется \emph{правило} или \emph{лемма Ито}.

Пусть $f := f(x^{k},t)$ --- функция от случайного процесса $x^{k}(t)$,
$f\in\mathrm{C}^{2}$. Тогда формула дифференциала будет выглядеть
следующим образом~\cite{sdu_10}:
\begin{equation}
  \label{eq:lemma_ito}
  \d f = \left[\partial_{t}f + a^{i}f_{,i} +
    \frac{1}{2}b^{i}_{a}b^{i a}f_{,ij}\right]\d t +
  b_{a}^{i}f_{,i}\d W^{a},
\end{equation}
где $f:=f(x^{k},t)$, $a^{i}:=a^{i}(x^{k},t)$,
$b^{i}_{a}:=b^{i}_{a}(x^{k},t)$ и $\d W^{a} := \d W^{a}(t)$.

\section{Модель <<хищник-жертва>>}
\label{sec:pp-model}

\subsection{Детерминистическая модель <<хищник-жертва>>}
\label{sec:pp-model:det}

Системы с взаимодействием двух видов популяций типа <<хищник-жертва>>
широко исследованы и для таких систем существует большое количество
разнообразных моделей. Самой первой моделью <<хищник-жертва>> принято
считать модель, полученную независимо друг от друга А.~Лоткой и
В.~Вольтеррой. Лотка в своей работе описывал некоторую
гипотетическую химическую реакцию~\cite{lotka}:
\begin{equation} 
\label{eq:0_4} 
A \xrightarrow{k_1}X \xrightarrow{k_2}
  Y \xrightarrow{k_3} B
\end{equation}
где $X,Y$ --- промежуточные вещества, коэффициенты $k_1,k_2,k_3$ ---
скорости химических реакций, $A$ --- исходный реагент, а $B$ ---
продукт реакции. В результате была получена система дифференциальных
уравнений вида:
\begin{equation} 
  \label{eq:0_5} 
  \begin{cases}
    \Dot{x} = k_1 x - k_{2} x y, \\
    \Dot{y} = k_{2} x y - k_3 y.
  \end{cases}
\end{equation}

Эта система полностью совпадает с системой дифференциальных уравнений,
полученной Вольтеррой, который рассматривал механизм роста численности
двух популяций с взаимодействием типа <<хищник-жертва>>. Для получения
уравнений Вольтерра делает ряд идеализированных предположений о
характере внутривидовых и межвидовых отношений в системе
<<хищник-жертва>>~\cite{p_5}.

\subsection{Стохастическая модель <<хищник-жертва>>}
\label{sec:pp-model:stoch}

Рассмотрим модель системы <<хищник--жертва>>, состоящую из особей двух
видов, причём один из них охотится, второй~--- обеспечен
неисчерпаемыми пищевыми ресурсами. Введя обозначения $X$~--- жертва,
$Y$~--- хищник, можно записать возможные процессы~\eqref{eq:chemkin2}
для вектора состояния $x^{\crd{i}} =
(X,Y)^T$~\cite{L_lit09,L_lit11,kulyabov:2013:conference:mephi,L_lit12}:
\begin{equation} 
  \label{eq:3_0}
  \begin{aligned}
    X &\ \xrightarrow{k_1} 2X, &\ r^{\crd{i} 1}=(1,0)^T \\
    X + Y &\ \xrightarrow{k_2} 2Y, &\ r^{\crd{i} 2} = (-1,1)^T \\
    Y &\ \xrightarrow{k_3} 0, &\ r^{\crd{i} 3}=(0,-1)^T,
  \end{aligned}
\end{equation}
которые имеют следующую интерпретацию. Первое соотношение означает,
что жертва, которая съедает единицу пищи, немедленно
репродуцируется. Второе соотношение описывает поглощение жертвы
хищником и мгновенное репродуцирование хищника. Это единственная
рассматриваемая возможность гибели жертвы. Последнее соотношение ---
это естественная смерть хищника.

Все процессы необратимы, поэтому $s^{-}_{\alpha}=0$, а
\begin{equation}
  \begin{gathered}
    s^+_1(x,y)= k_1 \frac{x!}{(x-1)!}\frac{y!}{y!}=k_1 x, \\
    s^+_2(x,y)= k_2\frac{x!}{(x-1)!}\frac{y!}{(y-1)!}=k_2 x y, \\
    s^+_3(x,y)= k_3\frac{x!}{x!}\frac{y!}{(y-1)!}=k_3 y.
  \end{gathered}
\end{equation} 

Воспользовавшись формулой~\eqref{eq:FP} имеем уравнение
Фоккера--Планка:
\begin{equation} 
  \label{eq:FP_hzh} 
  \frac{\partial p(x,y)}{\partial t} = 
  -\partial_i \left( A^{i} (x,y) p(x,y) \right) +
  \frac{1}{2} \partial_i\partial_j\left(B^{ij}(x,y) p(x,y)\right),
\end{equation}
где
\begin{equation}
  \begin{gathered}
    A^i(x,y) = s_{\alpha}^+ (x,y)\ r^{i \alpha},\\
    B^{ij}(x,y) = s_{\alpha}^+ (x,y) r^{i \alpha} r^{j \alpha}.
  \end{gathered}
\end{equation}

Таким образом мы получили:
\begin{equation}
\label{eq:pp_FP}
  \begin{multlined}
    A^{\crd{i}}(x,y)=
    \begin{pmatrix}
      1\\
      0
    \end{pmatrix}
    k_1 x +
    \begin{pmatrix}
      -1\\
      1
    \end{pmatrix}
    k_2xy + {} \\
                {} + 
    \begin{pmatrix}
      0\\
      -1
    \end{pmatrix}
    k_3y =
    \begin{pmatrix}
      k_1 x-k_2xy\\
      k_2xy - k_3y
    \end{pmatrix}, \\
      B^{\crd{i} \crd{j}}(x,y) =
      \begin{pmatrix}
        1\\
        0
      \end{pmatrix}
      (1,0) k_1 x +
      \begin{pmatrix}
        -1\\
        1
      \end{pmatrix}
      (-1,1) k_2xy + {} \\
                        {} + 
      \begin{pmatrix}
        0\\
        -1
      \end{pmatrix}
      (0,-1) k_3y  =
      \begin{pmatrix}
        k_1 x+k_2xy & -k_2xy \\
        -k_2xy & k_2xy + k_3y
      \end{pmatrix}.
  \end{multlined}
\end{equation}

Для того чтобы записать стохастическое дифференциальное уравнение в
форме Ланжевена~\eqref{eq:langevin} для модели <<хищник--жертва>>,
достаточно извлечь квадратный корень из полученной матрицы $B^{ij}$ в
уравнении Фоккера--Планка.
\begin{equation}
  \label{eq:pp_langevin}
  \begin{gathered}
  \d 
  \begin{pmatrix}
    x \\
    y
  \end{pmatrix}
  =
  \begin{pmatrix}
    k_1 x-k_2xy\\
    k_2xy - k_3y
  \end{pmatrix}\d t + 
        b^{\crd{i}}_{\crd{a}} 
  \begin{pmatrix}
    \d W^{1} \\
    \d W^{2}
  \end{pmatrix},\\
  b^{\crd{i}}_{a} b^{\crd{j} a} = B^{\crd{i}\crd{j}} =
    \begin{pmatrix}
      k_1 x+k_2xy & -k_2xy \\
      -k_2xy & k_2xy + k_3y
    \end{pmatrix}.
        \end{gathered}
\end{equation}

Следует заметить, что конкретный вид матрицы $b^{i}_{a}$ не выписан
из-за крайней громоздкости выражения. Впрочем, при дальнейших
исследованиях нам понадобится не собственно матрица $b^{i}_{a}$, а её
квадрат, то есть матрица $B^{ij}$.

\section{Реализация модели одношаговых стохастических процессов в системе компьютерной алгебры}
\label{sec:axiom}

\subsection{Обоснование выбора системы компьютерной алгебры}
\label{sec:axiom:compare}

Рассмотрим системы аналитических вычислений Maxima и Axiom.
Maxima является самой первой системой аналитических вычислений, 
написана она на языке Lisp. Maxima успешно работает на всех современных 
операционных системах: Windows, Linux и UNIX, Mac OS и даже на КПК 
под управлением Windows CE/Mobile. Документация интегрирована в программу в 
виде справочника, оснащенного средствами поиска. В Maxima нет разделения на 
объекты и данные, нет четкого разграничения между оператором и функцией. 
В системе нет встроенной графической отрисовки.

В отличие от Maxima, язык Axiom строго типизирован для лучшего отображения 
математических объектов и взаимосвязей. Вся математическая база написана 
на языке Spad. Переносимость Axiom чуть хуже: система работает под Linux, 
UNIX, а под Windows не работает построение графиков. Axiom обладает собственной 
графической подсистемой.

В 2007 году у Axiom появилось два форка с открытым исходным кодом: OpenAxiom и 
FriCAS. Open Axiom разрабатывается, придерживаясь идеологии Axiom, устраняются 
проблемы, которые встречались в Axiom.
Разработчики проекта FriCAS реорганизовали процесс сборки, расширили функционал.
Кроме того, FriCAS поддерживает не только GCL, который работает на ограниченном 
количестве платформ, но и ECL, Clisp, sbcl или openmcl, 
что позволяет запускать FriCAS на более широком множестве платформ.

\subsection{Описание реализации в системе компьютерной алгебры Axiom}
\label{sec:axiom:realization}

Метод стохастизации одношаговых процессов организован в виде модуля 
для системы компьютерной алгебры FriCAS. 
Для вывода всех вычислений на экран использована переменная \verb|SHOWCALC := true|.
Для вызова метода нужно воспользоваться основной функцией, которая имеет следующий 
вид:
\begin{verbatim}
osp(Matrix(Integer), Matrix(Integer), 
        Vector, Vector, Vector)
\end{verbatim}
где первый аргумент --- матрица $n^{i}_{j}$ состояний до взаимодействия, 
второй аргумент --- матрица $m^{i}_{j}$ после взаимодействия, 
третий аргумент --- вектор $k^{+}_{\alpha}$, 
четвертый аргумент --- вектор $k^{-}_{\alpha}$, 
пятый аргумент --- вектор состояний $x^{i}$.
Рассмотрим особенности языка FriCAS на вспомогательных функциях.
Например, функция \verb|calcProd| используется для упрощения вычислений 
$s^{+}_{\alpha}$ и $s^{-}_{\alpha}$. В реализации функции используется 
оператор условия и встроенная функция reduce:
\begin{verbatim}
calcProd : (Matrix(Integer), Vector, Integer, 
        Integer) -> Void
calcProd (n, x, a, i) == 
        nai:Integer := n(a,i) 
        if nai = 0 then 1 else reduce(*,
                [x(i) - j for j in 0..(nai-1)])
\end{verbatim}

В функции \verb|Bi| производятся промежуточные вычисления для 
элементов матрицы $B^{ij}$:
\begin{verbatim}
Bi (rv, sp, sm, i) == rv(i) * 
        (transpose rv(i)) * (sp(i) + sm(i))
\end{verbatim}

Чтобы воспользоваться модулем для расчёта модели системы <<хищник--жертва>>, 
вызываем функцию со следующими аргументами:
\begin{verbatim}
osp ([[1,0],[1,1],[0,1]],[[2,0],[0,2],[0,0]],
        vector([k1,k2,k3]), vector([0,0,0]),
               vector([x,y]))
\end{verbatim}

На рис.~\ref{fig:texmacs:pp_FP} представлен результат, получающийся в
оболочке \TeX{}macs. Фактически, мы повторили результаты, полученные
в~\eqref{eq:pp_FP}.

\begin{figure}
   \centering
   \includegraphics[width=0.9\linewidth]{matrix}
   \caption{Результат работы модуля для модели <<хищник-жертва>> в
     графической оболочке \TeX{}macs}
\label{fig:texmacs:pp_FP}
 \end{figure}

\section{Численный эксперимент в рамках программного комплекса}
\label{sec:numerical}

\subsection{Стохастические методы Рунге--Кутты}

Одним из хорошо известных численных методов решения СДУ является метод
Эйлера-Маруямы, который, является частным случаем более общих
стохастических методов Рунге--Кутты. Классический метод Рунге-Кутты
может быть обобщен на случай системы СДУ~\eqref{eq:langevin} следующим
образом~\cite{L_lit01, L_lit04}:
\begin{equation}
  \left\{
    \begin{aligned}
      X^{i}_{k} = & x_{0}^{i} +
      hR^{\crd{l}}_{k}a^{i}(X^{1}_{\crd{l}},\ldots,X^{n}_{\crd{l}}) +
      \widehat{R}^{\crd{l}}_{k}J^{\alpha}b^{i}_{\alpha}(X^{1}_{\crd{l}},\ldots,X^{n}_{\crd{l}}),\\
      x^{i}_{1} = & x_{0}^{i} +
      hr^{\crd{l}}a^{i}(X^{1}_{\crd{l}},\ldots,X^{n}_{\crd{l}}) +
      \hat{r}^{\crd{l}}J^{\alpha}b^{i}_{\alpha}(X^{1}_{\crd{l}},\ldots,X^{n}_{\crd{l}})
    \end{aligned}
  \right.
\end{equation}

Индексы $k = 1,\ldots,s$ и $l = 1,\ldots,n$ относятся к
стохастическому методу Рунге-Кутты. $J \sim N(0,h)$ или $J \sim
\sqrt{h}\varepsilon$, $\varepsilon \sim N(0,1)$ --- нормально
распределенные случайные величины. Выбор таких величин для численной
аппроксимации объясняется тем, что винеровские процесс реализуется как
$\d W = \varepsilon \sqrt{\d t}$. Также следует обратить внимание на
двойное суммирование в третьем слагаемом обеих формул численной схемы,
а также на то, что каждое число $J^{1},\ldots,J^{n}$ должно
генерироваться отдельно.

Коэффициенты метода, также как и для классического аналога, можно
сгруппировать в таблицу, называемую \emph{таблицей Батчера}:

\begin{equation}
\setstretch{1.8}
\begin{array}{c|c|c}
  &R_{ij}&        \hat{R}_{ij}\\ \hline
  &r_{j}&         \hat{r}_{j}
\end{array}.
\end{equation}

Для вычислений нами был использован метод с таблицей

\begin{equation}
\setstretch{1.2}
\begin{array}{c|ccc|ccc}
  &0&     0&      0&      0&      0&      0\\
  &2/3&   0&      0&      2/3&    0&      0\\
  &-1&    1&      0&      -1&     1&      0\\ \hline
  & 0&    3/4&    1/4&    0&      3/4&    1/4
\end{array}
\end{equation}

\subsection{Описание программной реализации} 

При написании комплекса программ ставилась задача как автоматизации 
вычисления коэффициентов $A^{i}$ и $B^{ij}$ СДУ с помощью описываемых 
выше общих принципов, так и численное решение полученного уравнения 
с помощью стохастических методов Рунге--Кутты. С точки зрения 
программирования задача распалась на три подзадачи:
\begin{enumerate}
\item генерация коэффициентов $A^{i}$ и $B^{ij}$ при помощи системы
  компьютерной алгебры;
\item генерация исходных текстов программ на языках
  Fortran и Julia, реализующих СДУ на основе полученных коэффициентов, 
  сохраненных в виде текстового файла;
\item написание подпрограмм/функций, реализующих стохастические методы
  Рунге-Кутты на языках Fortran и Julia, и дальнейшая компиляция их
  вместе с автоматически сгенерированными исходными текстами.
\end{enumerate}

Модуль Axiom в результате своей работы создает текстовой файл, в
котором содержатся коэффициенты $A^{i}$ и $B^{ij}$ в следующем виде:
\begin{verbatim}
        # A
        A[1]
        ...
        A[N]
        # B
        B[1,1] B[1,2] .. B[1,N]
        ...
        B[N,1] B[N,2] .. B[N,N] 
\end{verbatim} 
Матрица $b^{i}_{\alpha} = \sqrt{b^{i}_{\alpha}b^{j\alpha}} =
\sqrt{B^{ij}}$ вычисляется численно при помощи сингулярного разложения
матрицы (использована подпрограмма DGESVD библиотеки LAPACK).

Для решения второй подзадачи был выбран интерпретируемый язык Python
(версии 3).  Данный язык обладает широким набором средств для работы
со строками и текстовыми файлами. Кроме матриц $A^{i}$ и $B^{ij}$ была
задана дополнительная информация о математической модели в виде
словаря (стандартный тип данных в Python), где были указанны: название
модели, список переменных, список параметров, начальные значения
переменных, значения параметров и параметры численного метода (отрезок
интегрирования и величина шага).

На основе этих данных сценарий автоматически генерирует два файла
\verb|functions.f90| и \verb|main.f90|, где первый --- это модуль,
содержащий функции, задающие СДУ, а второй --- основной файл
программы. При компиляции к этим файлам добавляются еще три модуля со
вспомогательными процедурами, в числе которых содержится
стохастический метод Рунге-Кутты.

\subsection{Описание численного эксперимента}

В качестве модели для верификации работы комплекса программ
 была выбрана хорошо известная модель <<хищник-жертва>>, задаваемая 
 вектором $a^{i}$ с компонентами
\begin{equation}
  a^{1} = \alpha x - \beta xy,\;\; a^{2} = -\gamma y + \delta xy
\end{equation}
и матрицей $B^{ij}$:
\begin{equation}
  \left[
    \begin{array}{cc}
      \alpha x + \beta xy & -\beta xy \\
      -\beta xy & \beta xy + \gamma x
    \end{array}
  \right],
\end{equation}
$x$ --- число жертв, $y$ --- число хищников. Коэффициенты же имеют следующий физический (биологический) смысл:
$\alpha$ --- интенсивность роста популяции жертв,
$\beta$ --- частота встречи хищников и жертв,
$\gamma$ --- интенсивность смерти или миграции хищников в условии недостатка жертв,
$\delta$ --- интенсивность роста популяции хищников при условии избытка жертв.

При численном моделировании учитывалось, что значение переменных $x, y$ не может
быть меньше нуля, и при обращении в ноль одной из переменных работа программы 
прекращалась.

Численное моделирование показало, что добавление стохастики в классическую модель 
<<хищник-жертва>> приводит к тому, что по прошествии определенного времени наступает
 смерть одного из конкурирующих видов. Так, при следующих значениях параметров: 
 $\alpha = 10$, $\beta = 1,5$, $\gamma = 8,5$, $\delta = 1,8$ и следующих начальных 
 значениях: $x = 9,7$, $x = 6,77$ первыми  погибают жертвы, а за ними и 
 хищники ввиду недостатка пищи. Данный случай проиллюстрирован на рис.~\ref{fig:prepre_1_2}.  
 Для сравнения на рис.~\ref{fig:prepre_1_2d} представлен график для
 детерминированного случая.

\begin{figure}
  \centering
  \includegraphics[width=0.7\linewidth]{img_1_2}
\caption{Стохастическая модель <<хищник-жертва>> --- смерть жертв}
\label{fig:prepre_1_2}
\end{figure}

\begin{figure}
  \centering
  \includegraphics[width=0.7\linewidth]{img_1_2d}
  \caption{Детерминированная модель <<хищник-жертва>>}
\label{fig:prepre_1_2d}
\end{figure}

 При других условиях ($\alpha = 10$, $\beta = 1,5$, 
 $\gamma = 8,5$, $\delta = 0,5$, $x = 22$, $y = 6,76$) 
 погибают хищники, а численность жертв стремительно возрастает, так как для 
 них модель предполагает бесконечный источник пищи. Графики для 
 данного случая изображены на рис.~\ref{fig:prepre_2_2}, а 
 на рис.~\ref{fig:prepre_2_2d} для сравнения показан детерминированный
 случай.

 \begin{figure}
   \centering
   \includegraphics[width=0.7\linewidth]{img_2_2}
  \caption{Стохастическая модель <<хищник-жертва>> --- смерть
    хищников}
\label{fig:prepre_2_2}
\end{figure}

\begin{figure}
  \centering
  \includegraphics[width=0.7\linewidth]{img_2_2d}
  \caption{Детерминированная модель <<хищник-жертва>>}
\label{fig:prepre_2_2d}
\end{figure}

\section{Заключение}
\label{sec:conclusion}

В работе продемонстрировано применение разработанного метода формализации 
исходной физической системы, представимой в виде одного или нескольких 
одношаговых процессов. Формализация системы производится путем введения оператора эволюции. 
При этом для получения аналитического описания модели требуется применение большого числа 
рутинных операций. Для упрощения работы  предложено использовать средства компьютерной 
алгебры --- клон Axiom FriCAS.

Нами был  разработан аналитический программный блок пакета, получающий 
на входе  оператор эволюции, а на выходе выдающий систему СДУ, описывающую 
исходную модель. Для численного исследования полученной таким образом 
системы СДУ  разработан второй программный блок, который преобразует полученную 
в FriCAS систему уравнений в программный код на языке fortran и на выходе дает 
ее численное  решение. Таким образом, разработанный программный комплекс 
позволяет провести одновременно аналитическое и численное исследование исходной модели.

На данный момент разработанный программный комплекс охватывает не все возможности, 
заложенные в предложенном методе формализации исходной физической системы. 
Так в случае описания исходной системы с помощью ОДУ, граничные условия 
приходится вводить с помощью связей или индикаторных функций. 
В тоже время применение дифференциальных уравнений в частных производных 
может позволить решить эту проблему. Дальнейшие задачи --- разработка  полного 
программного комплекса применения  методики построения одношаговых моделий исходной 
физической системы.

\bibliographystyle{gost2008}

\bibliography{bib/axiom-sdu/axiom-sdu}